# Behavioral Targeting, a European Legal Perspective

Frederik Zuiderveen Borgesius
F.J.ZuiderveenBorgesius[at]uva.nl  Frederikzb[at]cs.ru.nl

Behavioral targeting, or online profiling, is a hotly debated topic. Much of the collection of personal information on the internet is related to behavioral targeting, while research suggests that most people don't want to receive behaviorally targeted advertising. The World Wide Web consortium is discussing a Do Not Track standard, and regulators worldwide are struggling to come up with answers. This article discusses European law and recent policy developments on behavioral targeting.

Behavioral targeting is the monitoring of people's online behavior over time to use the collected information to target people with advertising matching their inferred interests. In a simplified example, a company might assume that an internet user who often visits websites about recipes is a food enthusiast. If the user visits a news website, the user might see advertising for restaurants or cookbooks. When visiting that same news website, somebody who reads many blogs about computer science might see advertising for computer science books. Various technologies can be used for behavioral targeting, such as cookies, 'super cookies', device fingerprinting and deep packet inspection. Therefore, deleting cookies isn't always enough to prevent being tracked.[1]

---

[1] J. R. Mayer & J. C. Mitchell, "Third-Party Web Tracking: Policy and Technology", IEEE Security & Privacy, November/December 2012.

**Behavioral Targeting and European Law**

European data protection law, or privacy law, applies to behavioral targeting in most cases. In Europe the right to privacy and the right to data protection are fundamental rights. The European Convention on Human Rights (1953) protects the right to private life. The right to data protection is enshrined in the Charter of Fundamental Rights of the European Union (2009).

> Article 8 of the Charter of Fundamental Rights of the European Union
>
> Protection of personal data
>
> 1. Everyone has the right to the protection of personal data concerning him or her.
>
> 2. Such data must be processed fairly for specified purposes and on the basis of the consent of the person concerned or some other legitimate basis laid down by law. Everyone has the right of access to data which has been collected concerning him or her, and the right to have it rectified.
>
> 3. Compliance with these rules shall be subject to control by an independent authority.

The 1995 Data Protection Directive requires member states of the European Union to implement detailed data protection laws. People whose data are being processed have several rights, and parties that process personal data have a number of obligations. Independent Data Protection Authorities oversee compliance with the rules.

European data protection law is triggered when a company processes 'personal data'. Many behavioral targeting companies process pseudonymous profiles: individual but nameless profiles. Do these companies process 'personal data'? They usually do, say European Data Protection Authorities.

The Data Protection Directive defines personal data as: 'any information relating to an identified or identifiable natural person ("data subject").' A person is identifiable when he or she can be directly or indirectly identified. The Directive's preamble says: 'to determine whether a person is identifiable, account should be taken of all the means likely reasonably to be used either by the [company] or by any other person to identify the said person.'[2] Hence, it's not decisive whether it's the company holding the data, or another party that can identify a person. The definition of personal data is broader than the American concept of 'personally identifiable information'. But the exact scope of the definition has to be set by the courts.

The Court of Justice of the European Union, the highest authority on the interpretation of European Union law, hasn't ruled on behavioral targeting yet. But there is relevant case law. The discussion about behavioral targeting is similar to the debate about IP addresses. In a 2011 decision about IP addresses in the hands of an internet access provider, the Court said that those IP addresses were personal data. The Court thus confirms that information without a name can be personal data.[3]

---

[2] Article 2(a) and recital 26 of the Data Protection Directive 95/46.

European national Data Protection Authorities, cooperating in the Article 29 Working Party, say that data that can distinguish a person within a group are personal data. The Working Party adds that cookies used for behavioral targeting are personal data because they 'enable data subjects to be "singled out", even if their real names are not known'.[4] Although not legally binding, the Working Party's opinions are influential. Judges and national Data Protection Authorities often follow its interpretation.

In the Netherlands, the law presumes that behavioral targeting entails the processing of personal data. It would be up to the company to prove it doesn't process personal data.[5] Many scholars agree that data protection law usually applies to behavioral targeting.[6] Taking all this into account, it seems safe to assume that data protection law applies to behavioral targeting in most cases.

In January 2012, the European Commission presented proposals for a Data Protection Regulation that should replace the Directive from 1995. The proposed definition of personal data includes 'online identifiers' in the list of examples that may be used to identify a person.[7] Much discussion ensued. Many companies say that the law shouldn't apply to pseudonymous data. It could take some years before the Regulation is passed, and the text might still change.

**Fair Information Principles and European Law**

At the core of the European data protection regime are the fair information principles, or FIPs. Although the implementation varies from country to country, the fair information principles express a nearly worldwide consensus on how to ensure fair data processing. One of the first documents containing principles was a 1973 report by an advisory body to the government of the United States.[8] The American 1974 Privacy Act, which applies to government record systems, is based on the principles.

The fair information principles form the basis for international legal instruments, such as the OECD Data Processing Guidelines, and the Data Protection Convention, a treaty ratified by 44 4countries. Around 90 countries have a data privacy law based on the principles.[9] In 2012 reports on behavioral targeting, the Federal Trade Commission and the White House called for legislation that builds on the principles.

---

[3] Court of Justice of the European Union, 24 November 2011, Case C70/10 (Scarlet/Sabam), par. 51.
[4] Article 29 Working Party, "Opinion 2/2010 on Online Behavioral Advertising", (WP 171) 22 June 2010.
[5] Article 11.7a of the Telecommunications Act. See for an English translation of the clause: www.ivir.nl/publications/borgesius/Position_paper_W3C.pdf, p. 5.
[6] Behavioral targeting entails the processing of personal data: see e.g. P. Traung, "EU Law on Spyware, Web Bugs, Cookies, etc., Revisited: Article 5 of the Directive on Privacy and Electronic Communications", Business Law Review 2010-31, p. 216–228. See for the opposite view e.g. Interactive Advertising Bureau, "Isn't this a threat to my privacy?", Your Online Choices FAQ , www.youronlinechoices.com/uk/faqs#5.
[7] Proposal for a Regulation on the Protection of Individuals with Regard to the Processing of Personal Data and on the Free Movement of Such Data, COM(2012) 11 final.
[8] See for a concise overview: B. Gellman, "Fair Information Practices: A Basic History", http://bobgellman.com/rg-docs/rg-FIPShistory.pdf.

Now we turn to the principles as implemented in European law. Most important is the transparency principle. Secret data collection isn't allowed. A company must provide people whose data it processes all information that is needed to guarantee fair processing. The data quality principle requires companies to take reasonable steps to ensure they erase or rectify inaccurate data. It follows from the data minimization principle that parties shouldn't collect excessive amounts of data. The security principle requires an appropriate level of security for databases.

The purpose limitation principle says that data that are collected for one goal shouldn't be used for incompatible purposes. The purpose specification principle requires that personal data be collected for specified, explicit and legitimate purposes. A company needs a legitimate basis to process personal data. Personal data may be processed on the basis of the consent of the person concerned or some other basis laid down by law. In principle, three grounds could be relevant for the private sector: consent, a contract, or a legitimate business interest that overrides the fundamental rights of the data subject. But since 2009, European law always requires consent for behavioral targeting, as detailed below.

A company may process personal data for behavioral targeting if people give their unambiguous consent. Consent is defined as 'any freely given specific and informed indication of his wishes by which the data subject signifies his agreement to personal data relating to him being processed.' European Data Protection Authorities have elaborated on the requirements for consent.[10] Consent must be freely given, so consent given under pressure isn't valid. As consent has to be specific, consent 'to use personal data for commercial purposes' wouldn't be acceptable. In line with the transparency principle, consent has to be informed. Companies shouldn't hide relevant information in the fine print of a privacy policy. Consent can be given implicitly, but inactivity is almost never an indication of one's wishes. The proposed Data Protection Regulation further tightens the requirements for consent.

**European Law on Tracking Technologies**

The European Union has a separate directive for the protection of privacy and personal data in the electronic communications sector. Since 2009, this e-Privacy Directive requires companies to obtain consent of the internet user before they store or access cookies on a device.[11] (For ease of reading this article speaks of cookies, but the rule applies to many technologies, such as spyware and viruses.) No consent is needed for certain functional cookies, such as cookies used for log-in procedures or shopping carts. Regulators give a pop-up window with a choice to accept or reject cookies as an example of a proper way to ask consent.

---

[9] G. Greenleaf, "Global Data Privacy Laws: 89 Countries, and Accelerating", Privacy Laws & Business International Report, Issue 115, Special Supplement, February 2012; Queen Mary School of Law Legal Studies Research Paper No. 98/2012. Available at SSRN: http://ssrn.com/abstract=2000034.

[10] Article 29 Working Party, "Opinion 15/2011 on the Definition of Consent", (WP 187) 13 July 2011.

[11] Article 5.3 and recital 17 of the e-Privacy Directive 2002/58, amended by Directive 2009/136; recital 66 of Directive 2009/136.

The e-Privacy Directive's preamble says that people can express their consent 'by using the appropriate settings of a browser or other application'. But it refers to the definition of consent in the Data Protection Directive: a free, specific and informed indication of will. Some say that *default* browser settings can express consent for tracking cookies. Others say that people that don't tweak their browser may be unaware of accepting tracking cookies. Therefore there's no consent. These opposite views have led to much debate.

It's unclear how the e-Privacy Directive will be implemented in all the European Union member states. The approaches seem to vary. For instance, the Netherlands requires explicit consent for tracking cookies. But other countries appear to apply a lower threshold. The United Kingdom seems to accept some kind of implicit consent if websites give sufficient notice.[12] In short, the e-Privacy Directive requires companies to obtain informed consent for tracking technologies, but it's contested how consent should be obtained.

To whom do the rules apply? Website publishers and behavioral targeting companies in Europe have to comply. Many American companies, such as Facebook and Google, are based in Ireland, and must abide as well. Moreover, European Data Protection Authorities say that the rules apply to any company that stores a cookie on a device in Europe.

**Do Not Track**

Since September 2011, a Tracking Protection Working Group of the World Wide Web Consortium has been engaged in a discussion about a Do Not Track standard. The standard would allow people to signal with their browser that they don't want to be tracked. A company that receives a 'Do not track me' signal could reply: 'OK, I won't track you'.

But if the company continued to track the user after such a reply, the law might come into play. In the United States for instance, the Federal Trade Commission (FTC) can intervene when companies break their promises.

Neelie Kroes, a member of the European Commission, has suggested that a Do Not Track system could make it possible for companies to comply with the e-Privacy Directive's consent requirement.[13] It's not evident how Do Not Track – an opt-out system – could help companies to comply with the e-Privacy Directive. But perhaps an arrangement among the following lines would be possible.

Companies should refrain from tracking European internet users that haven't set a Do Not Track preference. If somebody signals to a company 'Yes, track me' after receiving sufficient information, that company may place a tracking cookie. Hence, in Europe not setting a preference would have the same effect as setting a preference for 'Do not track me'. In the United States, the law doesn't require consent, so companies might be free to track people that don't set a preference.

---

[12] Information Commissioner's Office of the United Kingdom, "Guidance on the Rules on Use of Cookies and Similar Technologies (V.3)", May 2012, www.ico.gov.uk/for_organisations/privacy_and_electronic_communications/the_guide/cookies.aspx. The Netherlands: article 11.7a of the Telecommunications Act. See for an English translation of the clause: www.ivir.nl/publications/borgesius/Position_paper_W3C.pdf, p. 5.

[13] N. Kroes, Speech at Online Tracking Protection & Browsers Workshop, Brussels 22 June 2011, http://europa.eu/rapid/press-release_SPEECH-11-461_en.htm.

The proposals for a Do Not Track standard exclude tracking within one website. Therefore, the standard would allow companies like Amazon or Facebook to analyze people's behavior within their own website, even if people signal 'Do not track me'. In contrast, the e-Privacy Directive's consent rule also applies to first party tracking cookies.

Many issues still have to be decided regarding the standard. One of the main points of disagreement is whether Do Not Track means 'Do Not Collect' (not use tracking technologies) or merely 'Do Not Target' (continue collecting data but stop showing targeted advertising). Another dispute is whether a browser that signals 'Do not track me' by default should be respected. Microsoft sets its Internet Explorer 10 to 'Do not track me' by default. Some advertising companies say that such default signals can be ignored because they don't express a user's choice.[14] At the time of writing the Tracking Protection Working Group hasn't reached consensus.

**Fair Information Principles and Technological Innovation**

In conclusion, European data protection applies to behavioral targeting in many cases. Like most privacy laws in the world, the European law is centered around the fair information principles. While the national implementation varies considerably, there is nearly worldwide consensus on the gist of the principles.

But the law alone may not be enough to put the principles in practice. For example, privacy policies that nobody reads or understands do little to make data processing transparent. Data collection by websites and smart phone apps remains opaque for most people. Meaningful 'informed consent' is rare.

According to its mission statement, 'IEEE's core purpose is to foster technological innovation and excellence for the benefit of humanity.' Much work is being done on data security for safeguarding personal data. There is research on behavioral targeting systems that don't rely on sharing one's browsing behavior with others, thereby minimizing data collection.[15] But hopefully more progress is possible. Perhaps the fair information principles could provide inspiration. For example, might there be technological solutions for making data processing transparent, or to enable people to access data concerning them? Could technology foster fair information processing?

* * *

---

[14] See the discussion on the public mailing list of the Do Not Track Working Group of the World Wide Web consortium: http://lists.w3.org/Archives/Public/public-tracking/.

[15] S. Barocas, D. Boneh V. Toubiana, A. Narayanan & H. Nissenbaum, "Adnostic: Privacy Preserving Targeted Advertising", 2010, http://crypto.stanford.edu/adnostic.

**Acknowledgements**
This article is based on a position paper presented at the W3C Do Not Track and Beyond Workshop in Berkeley (26 and 27 November 2012), www.w3.org/2012/dnt-ws.